\documentstyle[preprint,epsfig,aps,eqsecnum,twoside]{revtex}
\newcommand{\C}{{\if mm {{\rm C}\mkern -15mu{\phantom{\rm t}\vrule}}
\mkern +10mu \else \leavemode \hbox{I}\kern -.17em \hbox{C} \fi}}
\newcommand {\vf}{\varphi}
\newcommand {\pr}{\partial}
\newcommand {\vt}{\vartheta}
\newcommand {\p}{\prime}

\begin{document}
\title{Scalar Field with Induced Nonlinearity: 
Regular Solutions and their Stability}
\author{B.~Saha \\
{\small \it Bogoliubov Laboratory of Theoretical Physics}\\
{\small \it Joint Institute for Nuclear Research \\
Dubna, 141980, Moscow reg., Russia\\
e-mail: saha@thsun1.jinr.dubna.su}}
\maketitle
\begin{abstract}
Exact particle-like static, spherically and/or cylindrically symmetric 
solutions to the equations of interacting scalar 
and electromagnetic field system have been obtained within the scope of
general relativity. In particular, we
considered Freedman-Robertson-Walker (FRW) space-time as  
an external homogenous and isotropic gravitational field
whereas the inhomogenous and isotropic Universe is given by the
G$\ddot o$del model. The solutions obtained have been thoroughly studied
for different types of interaction term. It has been shown that in FRW 
space-time equations with different interaction terms may have stable 
solutions while within the scope of G$\ddot o$del model only the 
droplet-like configurations may be stable, if they are located in the region 
where $g^{00}>0$.  
\end{abstract}
\vskip 3mm
\noindent
{\bf Key words:} Freedman-Robertson-Walker (RFW) space-time, droplets                        
\vskip 3mm
\noindent
{\bf PACS 04.20.Jb} Exact Solutions
\section{Introduction} 
\setcounter{equation}{0}
Development of general relativity (GR) and quantum field theory (QFT) 
leads to the increasing interest to study the role of gravitational 
field in elementary particle physics. To obtain and study the properties 
of regular localized solutions to the nonlinear classical field equations  
is motivated mainly by a hope to create a consistent, divergence-free theory. 
These solutions, as was remarked by Rajaraman~\cite{Rajaraman} give us one 
of the ways of modeling elementary particles as extended objects with 
complicated spatial structure. In such attempts it is natural to treat
the field nonlinearity not only as a tool for avoiding the theoretical
difficulties (such as singularities) but also as a reflection of real
properties of physical system. It should be also emphasized that the 
complete description of elementary particles with all their physical 
characteristics (e.g., magnetic momentum) can be given only in the 
framework of interacting field theory~\cite{Bogoliubov}. 
In this paper we present some regular particle-like solutions within
the scope of general relativity for an interacting system of scalar
and electromagnetic fields, confining ourselves to static, spherically
and/or cylindrically symmetric configurations.
\section{Fundamental equations} 
\setcounter{equation}{0}
We consider a system with the Lagrangian 
\begin{equation}
L = {{R}\over{2 \kappa}} + {{1}\over{2}} \vf_{,\alpha} \varphi^{,\alpha} 
- {{1}\over{16 \pi}} F_{\alpha \beta}{F}^{\alpha \beta} \Psi(\vf) 
\label{lag}
\end{equation} 
where the first term describes the Einstein gravitational field
($R$ is the scalar curvature, $\kappa = 8 \pi G$ is the gravitational
constant) and $\Psi(\vf)$ is some arbitrary function characterizing 
interaction between the scalar $(\vf)$ and electromagnetic 
$(F_{\mu\nu})$ fields. This kind of interaction has been thoroughly discussed
in~\cite{Staniukovich}. The nongravitational part of the Lagrangian(~\ref{lag})
describes a system with a positive-definite energy if and only if 
$\Psi \ge 0$.
Let us write the scalar and the electromagnetic field equations  
corresponding to the Lagrangian(~\ref{lag})
\begin{equation}
{{1}\over{\sqrt{-g}}} {{\pr}\over{\pr x^\nu}}\biggl( \sqrt{-g} g^{\nu\mu}
{{\pr \vf}\over{\pr x^\mu}}\biggr) +
{{1}\over{16 \pi}} F_{\alpha \beta}{F}^{\alpha \beta} \Psi_{\vf} = 0, \quad
\Psi_{\vf} = {{\pr \Psi}\over{\pr \vf}}
\label{scl}
\end{equation}
\begin{equation} 
{{1}\over{\sqrt{-g}}}
{{\pr}\over{\pr x^\nu}}\biggl(\sqrt{-g} F^{\nu\mu} \Psi (\vf) \biggr) = 0
\label{em}
\end{equation}
The corresponding energy-momentum tensor reads
\begin{equation} 
T_{\mu}^{\nu} = \vf_{,\mu}\varphi^{,\nu} - {{1}\over{4 \pi}}
F_{\mu\beta} F^{\nu \beta} \Psi(\vf) - \delta_{\mu}^{\nu}\Bigl[ 
{{1}\over{2}} \vf_{,\alpha} \varphi^{,\alpha} 
- {{1}\over{16 \pi}} F_{\alpha \beta}{F}^{\alpha \beta} \Psi(\vf) \Bigr]
\label{emt}
\end{equation} 
\section{Spherically symmetric configurations}
\setcounter{equation}{0}
\subsection{Solutions in FRW Universe}
As an external homogenous and isotropic gravitational field we 
choose the FRW space-time. This Universe is very important as 
the corresponding cosmological models coincides with observation. 
The interval in the FRW Universe in general takes the form
~\cite{Weinberg,Birrell}
\begin{equation}
ds^2 = dt^2 - R^2(t)\biggl[{{dr^2}\over{1 - kr^2}} + 
r^2 \bigl\{d\vt^2 + {\rm sin}^2\vt d\phi^2 \bigr\}\biggr] 
\label{met}
\end{equation}
Here $R(t)$ defines the size of the Universe, and $k$ 
takes the values $0$ and $\pm 1$. We consider the simple most case
putting $R(t)\,=\,R\,=$ constant, which corresponds to the static FRW
Universe. In static case $k = 0$ corresponds to usual Minkowski space,
$k = +1$ describes the close Einstein Universe \cite{Ein} and $k = -1$ 
corresponds to the space-time with hyperbolic spatial cross-section.
Note that the velocity of light $c$ has been taken to be unity.
 
As was mentioned earlier, we seek the static, spherically symmetric
solutions to the equations (~\ref{scl}) and (~\ref{em}). To this end
we assume that the scalar field is the function of $r$ only, 
i.e. $\vf = \vf (r)$ and the electromagnetic field possesses only one 
component $F_{10} = \pr A_0 / \pr r = A^{\p}$.
                   
Under the assumption made above, the solution to the equation (~\ref{em}) 
reads
\begin{equation} 
F^{01} = {\bar q} P(\vf) {{\sqrt{1 - k r^2}}\over{R^3 r^2}} \label{emf}
\end{equation} 
where ${\bar q}$ is the constant of integration and 
$P(\vf) = 1 / \Psi (\vf).$
Putting (~\ref{emf}) into (~\ref{scl}) for the scalar field we obtain
the equation with "induced nonlinearity" ~\cite{BronIz}  
\begin{equation}
(1 - k r^2) \vf^{\p \p} + {{2 - 3 k r^2}\over{r}} \vf^{\p} - 
{{2 q^2}\over{R^2 r^4}} P_{\vf} = 0, \quad q^2 = {{{\bar q}^2}\over{16 \pi}}
\label{sfd}
\end{equation}
This equation can be written in the form   
\begin{equation}
{{\pr^2 \vf}\over{\pr z^2}} - {{2 q^2}\over{R^2}} P_{\vf} = 0
\label{fe}
\end{equation}
with the first integral
\begin{equation}
{{\pr \vf}\over{\pr z}} = {{2 q}\over{R}} \sqrt{P + C_0}
\label{int1}
\end{equation}
where we substitute $z = \sqrt{1/r^2 - k}$.
Here $C_0$ is the constant of integration, which under the regularity 
condition of $T_{0}^{0}$ at the center turns to be trivial, i.e., $C_0 = 0$. 
Finally we write the solution to the scalar field equation in quadrature
\begin{equation}
\int {{\pr \vf}\over{\sqrt{P}}} =  {{2 q}\over{R}} (z - z_0)
\label{qua}
\end{equation}
In accordance with (~\ref{emf}) and (~\ref{int1}) from (~\ref{emt}) 
we find the density of field energy of the system 
\begin{equation} 
T_{0}^{0} = {{4 q^2 P}\over{R^4 r^4}} 
\label{eden}
\end{equation} 
and total energy of the material field system
\begin{equation} 
E_f = \int T_{0}^{0} \sqrt{-^3g} d^3{\bf x} = 
- 8 \pi q \int \sqrt{P}d\vf
\label{toten}
\end{equation} 
Thus, we see that the energy density $T_{0}^{0}$ and total energy $E_f$ of 
the configurations obtained do not depend on the conventional values of 
the parameter $k$. As one sees, to write the scalar $(\vf)$ 
and vector $(A)$ functions as well as the energy density $(T_{0}^{0})$
and energy of the material fields $(E_f)$ explicitly, one has to give  
$P(\vf)$ in explicit form. Here we will give a detailed analysis for
some concrete forms of $P(\vf)$.
Let us choose $P(\vf)$ in the form
\begin{equation}
P(\vf) = P_0 {\rm cos}^2 \bigl({{\lambda \vf}\over{2}}\bigr)
\label{cos}
\end{equation}
with $\lambda$ being the interaction parameter. Inserting~(\ref{cos}) into 
(~\ref{fe}) we get the sin-Gordon type equation~\cite{Shikin}
\begin{equation}
{{\pr^2 \vf}\over{\pr z^2}} + {{\lambda q^2 P_0}\over{4 R^2}} 
{\rm sin}(\lambda \vf) = 0
\label{sG}
\end{equation}
with the solution
\begin{equation}
\vf (z) = {{2}\over{\lambda}} {\rm arcsin\, tanh}[b (z + z_1)], \quad
b = {{\lambda q \sqrt{P_0}}\over{R}}, \quad z_1 = {\rm const} 
\label{ssG}
\end{equation}
Let us analyze the solution (~\ref{ssG}). It can be shown that
$\lim\limits_{r \to 0} \vf = \pi/\lambda$ for all $k$. As one sees,
the solution possesses meaning only in the region where
$z = \sqrt{1/r^2 - k} > 0$. It means, in case of $k = + 1$ the configuration
confines in the interval $0\le r \le 1$. Then the asymptotic behavior of
the solution (~\ref{ssG}) can be written as
\begin{equation}
\vf \rightarrow \left\{\begin{array}{ccc} 
0,&  k = -1, \quad z_1 = -1, \quad r \to \infty \\
0,&  k = 0, \quad z_1 = 0, \quad r \to \infty \\ 
0,&  k = + 1, \quad z_1=0, \quad r \ge 1
\end{array}\right.
\end{equation}
From (~\ref{toten}) we find the total energy of the system 
$E_f = - 16\pi q \sqrt{P_0} / \lambda$.
For the choice of $P(\vf)$ in the form
\begin{equation}
P = \lambda (a^2 - \vf^2)^2 
\label{kdv}
\end{equation}
with $\lambda$ being the coupling constant and $a$ being some arbitrary
constant, from (~\ref{fe}) we obtain
\begin{equation}
4 \lambda a^2 \vf - 4 \lambda \vf^3 + {{\pr^2 \vf}\over{\pr z^2}} = 0
\label{MKdV}
\end{equation}
The equation (\ref{MKdV}) can be seen as an MKdV one. Indeed, a KdV equation
\begin{equation}
{{\pr u}\over{\pr t}} + \alpha u^p {{\pr u}\over {\pr x}} + 
\beta {{\pr^3 u}\over{\pr x^3}} = 0,
\end{equation}
can always be converted to 
\begin{equation}
-Du + \alpha {{u^{p+1}}\over{p+1}} + \beta {{d^2 u}\over{d z^2}} = 0
\end{equation}
if one looks for stationary solution of the form $u = u(z)$ where
$z = x - D t$. In our particular case $p =2$ and the equation (\ref{MKdV})
is an MKdV one.  
The scalar field function in this case has the form
\begin{equation}
\vf (z) = a {\rm tanh} [\sqrt{\lambda} a b (z + z_2)], \quad b ={{2q}\over{R}}
\label{th}
\end{equation}
Taking into account that $z =\sqrt{1/r^2 -k}$ one sees that at the origin
$\lim\limits_{r \to 0} \vf = a$ whereas at the asymptotic region for
different value of $k$ we get
\begin{equation}
\vf \rightarrow \left\{\begin{array}{ccc} 
0,&  k = -1, \quad z_2 = -1, \quad r \to \infty \\
0,&  k = 0, \quad z_2 = 0, \quad r \to \infty \\ 
0,&  k = + 1, \quad z_2=0, \quad r \ge 1
\end{array}\right.
\end{equation}
From (~\ref{toten}) we find the total energy of the system to be
$E_f = - 16\pi q \sqrt{\lambda} a^3 /3$.
A specific type of solution to the nonlinear field equations in flat 
space-time was obtained in a series of interesting articles~\cite{Werle}. 
These solutions are known as droplet-like solutions or simply droplets. 
The distinguishing property of these solutions is the availability of some 
sharp boundary defining the space domain in which the material 
field happens to be located, i.e., the field is zero beyond this area. It 
was found that the solutions mentioned exist in field theory with specific 
interactions that can be considered as an effective one, generated by initial 
interactions of unknown origin. In contrast to  the  widely 
known soliton-like solutions, with field  functions  and  energy 
density asymptotically tending to zero at spatial infinity,  the 
solutions in question vanish at a finite distance from the center 
of the system (in the case of spherical  symmetry)  or  from  the 
axis (in the case of cylindrical symmetry).  Thus, there  exists 
a sphere or cylinder with critical radius $r_0$ outside of which 
the fields disappear. Therefore the field configurations have a 
droplet-like structure~\cite{BronIz,Werle,Rybakov}. 
  
To obtain the droplet-like configuration we choose a very specific type of 
interaction function $P(\varphi)$ which has the form~\cite{Saha} 
\begin{equation} 
P(\vf) =  J^{2-4/\sigma} \biggl(1 - J^{2/\sigma}\biggr)^2 \label{P}
\end{equation}
where $J = \lambda \vf, \quad \sigma = 2n + 1, \quad n =  
1,\, 2 \cdots$. Putting (~\ref{P}) into (~\ref{qua}) for $\vf$ one gets 
\begin{equation} 
\vf (z)  = {{1}\over{\lambda}} \biggl[1 - 
{\rm exp} \biggl(-{{4 q \lambda}\over {R \sigma}}
\bigl(z - z_0 \bigr)\biggr)\biggr]^{\sigma/2} \label{drop}
\end{equation} 
Recalling that $z = \sqrt{1/r^2 -k}$
from (~\ref{drop}) we see that at $r\to 0$ the scalar field $\vf$ takes
the value $\vf (0) \to 1/\lambda$ and at $r \to r_c = 1/\sqrt{z_0^2 + k}$, 
the scalar field function becomes trivial, i.e.,  
$\vf (r_c) \to 0.$ It is obvious that for $r > r_c$ 
the value of the square bracket turns out to be negative and $\vf (r)$ 
becomes imaginary, since $\sigma$ is an odd number. Since we are 
interested in real $\vf$ only, without loss of generality 
we may assume the value of $\vf$ to be zero for $r \ge r_c$, the
matching at $r = r_c$ (i.e., $z = z_0$) being smooth. Note that, for
$k = +1$, the scalar field is confined in the region $0\le r \le 1$, as
it was in previous two cases.
The total energy of the droplet we obtain from 
(~\ref{toten}) has the form
\begin{equation}
E_f = {{4 \pi q}\over{\lambda (\sigma - 1)}}
\label{te}
\end{equation}
As is seen from (~\ref{te}), the value of the total energy does not depend
on the size of the droplet, it means droplets of different linear size
share the same total energy.
\subsection{Stability problem}
To study the stability of the configurations obtained we write the 
linearized equations for the radial perturbations of scalar field 
assuming that 
\begin{equation} 
\vf (r, t) = \vf (r) + \xi (r, t), \quad \xi \ll \varphi 
\label{per}
\end{equation} 
Putting (~\ref{per}) into (~\ref{scl}) in view of (~\ref{sfd}) we get the 
equation for $\xi(r, t)$ 
\begin{equation} 
\ddot {\xi} + 3 {{\dot R}\over{R}} \dot \xi - {{1 - k r^2}\over{R^2}} 
\xi^{\p \p} - {{2 - 3 k r^2}\over{r R^2}} \xi^{\p} + 
{{q^2 P_{\vf \vf}}\over{R^4 r^4}} \xi = 0 \label{efp} 
\end{equation} 
The second term in (~\ref{efp}) is zero since we assume the FRW space-time
to be static one putting $R=$ constant. Assuming that 
\begin{equation} 
\xi(r, t) \approx  v(r) {\rm exp}(-i \Omega t), \quad 
\Omega = \omega/R \label{om}
\end{equation} 
from (~\ref{efp}) we obtain 
\begin{equation} 
(1 - k r^2) v^{\p \p} - {{2 - 3 k r^2}\over{r}} v^{\p} + 
\Bigl[\omega^2 - {{q^2 P_{\vf \vf}}\over{R^2 r^4}} \Bigr] v = 0 \label{efv} 
\end{equation} 
The substitution 
\begin{equation}
\eta (\zeta) = r \cdot v(r), \quad 
\zeta = {{1}\over{\sqrt{k}}}{\rm arcsin} (\sqrt{k} r)
\label{eta}
\end{equation}
leads the equation (~\ref{efv}) to the Liouville one~\cite{Kamke}  
\begin{equation} 
\eta_{\zeta\zeta} + 
\bigl(\omega^2 - V (\vf)\bigr) \eta = 0, \quad
V (\vf) = - k + {{q^2 P_{\vf \vf}}\over{R^2 \zeta^4}} 
\Bigl({{\sqrt{k}\zeta}\over{{\rm sin} (\sqrt{k}\zeta)}}\Bigr)^4 
\end{equation} 
For the interaction term (~\ref{cos}) we see that
\begin{equation}
V (\vf) >0,\quad {\rm if\,\, and\,\, only\,\, if} \quad
{\rm tanh}^2 b(z + z_1) > {{k\zeta^4 R^2}\over{P_0 q^2 \lambda^2}}
\Bigl({{{\rm sin} (\sqrt{k}\zeta)}\over{{\sqrt{k}\zeta}}}\Bigr)^4 
+{{1}\over{2}} \label{cosstab}
\end{equation}
Thus we find that the equations with trigonometric nonlinearity contain
stable solutions.
Given $P (\vf)$ in the form (~\ref{kdv}) we find that
\begin{equation}
V (\vf) > 0, \quad{\rm if} \quad {\rm tanh}^2 b(z + z_2) >
{{k\zeta^4 R^2}\over{12 q^2 \lambda}}
\Bigl({{{\rm sin} (\sqrt{k}\zeta)}\over{{\sqrt{k}\zeta}}}\Bigr)^4
+{{a^2}\over{3}} \label{kdvstab}
\end{equation}
As is seen from (~\ref{kdvstab}) the equations with polynomial type of 
nonlinearity too contain some stable solutions.
For the droplet-like configurations, i.e., for the interacting term 
$P(\vf)$ given by (~\ref{P}), it can be shown that the potential 
\begin{equation}
\lim\limits_{r \to 0} V (\vf) \to + \infty, \quad
\lim\limits_{r \to r_c} V (\vf) \to + \infty
\label{pot}
\end{equation} 
beginning with $\sigma \ge 5$. It means that the droplet-like 
configurations (~\ref{drop}) with $\sigma \ge 5$ are stable for the 
class of perturbation, vanishing at $r = 0$ and $r = r_c$.
 
\section{Cylindrically symmetric configurations}
\setcounter{equation}{0}
\subsection{Solutions in G$\ddot o$del Universe}
In the previous section we studied the possibility of formation of
regular localized configuration in homogenous and isotropic FRW Universe.
Let us now continue our study in the homogenous but anisotropic Universe. 
In particular we consider the model proposed by G$\ddot o$del. 
The linear element of G$\ddot o$del Universe in cylindrical coordinates 
reads~\cite{God}
\begin{equation}
ds^2 = dt^2 - d\rho^2 + {{1}\over{\Omega^2}}[{\rm sinh}^4 \Omega \rho
- {\rm sinh}^2 \Omega \rho] d \phi^2 -{{\sqrt{8}}\over{\Omega}}
{\rm sinh}^2 \Omega \rho d\phi dt - dz^2 \label{gmet}
\end{equation}
where the constant $\Omega$ is related with the angular velocity $\omega$:
$\omega = \sqrt{2}\Omega$. This form of linear element of the 
four-dimensional homogenous space $S$ directly exhibits its rotational
symmetry, since the $g_{\mu\nu}$ do not depend on $\phi$. It is easy to 
find  
\begin{eqnarray}
\lim\limits_{\Omega \to 0} \sqrt{-g} =
\lim\limits_{\Omega \to 0} {{1}\over{2\Omega}} {\rm sinh}(2 \Omega \rho)
\to \rho \nonumber
\end{eqnarray}
i.e. at $\omega \to 0$ G$\ddot o$del Universe transfers to the flat one.
 
As in spherically symmetric case, here too we seek the static 
solutions to the equations (~\ref{scl}) and (~\ref{em})
assuming the scalar field to be the function of $\rho$ only, i.e., 
$\vf = \vf (\rho)$ and the electromagnetic possesses only one component 
$F_{10} = \pr A_0 / \pr \rho = A^{\p}$.
For the electromagnetic field in this case we find
\begin{equation}
F^{01} = 2\Omega D P/{\rm sinh} (2\Omega\rho), \quad D = {\rm const}
\label{eg}
\end{equation}
The scalar field equation (~\ref{scl}) with regards to (~\ref{eg}) reads
\begin{equation}
{{\pr^2 \vf}\over{\pr \rho^2}} + 2 \Omega {\rm coth} (2 \Omega \rho)
{{\pr \vf}\over{\pr \rho}} =  
{{8 q^2 \Omega^2 P_\vf}\over{{\rm sinh}^2 (2 \Omega\rho)}}, 
\quad q^2 = D^2/ 16 \pi \label{sfg}
\end{equation}
Putting $y = {{1}\over{2 \Omega}} {\rm ln\,tanh} (\Omega\rho)$ from (~\ref{sfg})
one gets
\begin{equation}
{{\pr^2 \vf}\over{\pr y^2}} -  8 q^2 \Omega^2 P_\vf = 0
\label{sfeg}
\end{equation}
with the first integral 
\begin{equation}
{{\pr \vf}\over{\pr y}} = \pm 4 q \Omega \sqrt{P + D_0}
\label{int2}
\end{equation}
Here $D_0$ is the constant of integration, which under the regularity 
condition of $T_{0}^{0}$ at the center turns to be trivial, i.e., $D_0 = 0$. 
Finally we write the solution to the scalar field equation in quadrature
\begin{equation}
\int {{\pr \vf}\over{\sqrt{P}}} =  4 q \Omega (y - y_0)
\label{quag}
\end{equation}
In accordance with (~\ref{eg}) and (~\ref{int2}) from (~\ref{emt}) 
we find the density of field energy and the total energy of the system 
\begin{eqnarray} 
T_{0}^{0} &=& {{16 q^2 \Omega^2 P}\over{{\rm sinh}^2 (2\Omega\rho)}} 
\label{edeng}\\
E_f &=&  8 \pi q \int \sqrt{P}d\vf
\label{toteng}
\end{eqnarray} 
As in the previous case, to write the scalar $(\vf)$ 
and vector $(A)$ functions as well as the energy density $(T_{0}^{0})$
and energy of the material fields $(E_f)$ explicitly, one has to give  
$P(\vf)$ in explicit form. Here again we will thoroughly study the solutions 
obtained for different concrete forms of $P(\vf)$.
Choosing $P(\vf)$ in the form (~\ref{cos}), i.e.,
\begin{equation}
P(\vf) = P_0 {\rm cos}^2 \bigl({{\lambda \vf}\over{2}}\bigr)
\label{cos1}
\end{equation}
with $\lambda$ being the interaction parameter, from (~\ref{sfeg}) we 
get the sin-Gordon type equation
\begin{equation}
{{\pr^2 \vf}\over{\pr y^2}} + 4\lambda q^2 \Omega^2 P_0 
{\rm sin}(\lambda \vf) = 0
\label{sG1}
\end{equation}
The solution to this equation can be written in the form
\begin{equation}
\vf (\rho) = {{4}\over{\lambda}} \Biggl[{\rm arctan}
\biggl({{{\rm tanh} \Omega \rho}\over{\Omega\rho_0}}\biggr)^{\alpha} - 
{{\pi}\over{4}}\Biggr], \quad
\alpha = \lambda q \sqrt{P_0}
\label{ssG1}
\end{equation}
where $\rho_0$ is the constant of integration, giving the size of the system.
Without losing the generality we can choose $\alpha > 0$. Then one finds
$\lim\limits_{\rho \to 0} \vf = -\pi/\lambda$. For $\rho > 0$ the field 
$\vf$ steadily increases up to $\pi/\lambda$. In particular, at spatial 
infinity we get $\lim\limits_{\rho \to \infty} \vf = 0$. In this case 
$P(\vf_\infty) = 1$ which corresponds to the exclusion of interaction
at spatial infinity. The total energy of the system in this case coincides
with that of in FRW Universe, i.e., $E_f = - 16 \pi q \sqrt{P_0}/ \lambda$.
Choosing $P(\vf)$ in the form (~\ref{kdv}), i.e.,
\begin{equation}
P = \lambda (a^2 - \vf^2)^2 
\label{kdv1}
\end{equation}
from (~\ref{sfeg}) we as in FRW case again obtain MKdV type equation
with the solution
\begin{equation}
\vf (\rho) = a {{{\rm tanh}^\alpha (\Omega\rho) -1}
\over{{\rm tanh}^\alpha (\Omega\rho) +1}}, \quad \alpha = 4 a q \sqrt{\lambda}
\label{mkdvg}
\end{equation}
From (~\ref{mkdvg}) it is clear that $\lim\limits_{\rho \to 0} \vf \to - a$
and $\lim\limits_{\rho \to \infty} \vf \to  0$
From (~\ref{toteng}) we find the total energy of the system to be
$E_f = - 16\pi q \sqrt{\lambda} a^3 /3$ as it was in FRW case.
The choice of the interaction term $P(\vf)$ in the form (~\ref{P}), i.e.,
\begin{equation} 
P(\vf) = J^{2-4/\sigma}\bigl(1 - J^{2/\sigma}\bigr)^2 
\end{equation}
where $J = \lambda \vf, \quad \sigma = 2n+1, \quad n =  
1,\, 2 \cdots$, leads to the following expression for scalar field  
\begin{equation} 
\vf(\rho) = {{1}\over{\lambda}} \biggl[1 - 
\Bigl({{{\rm tanh}(\Omega\rho)}\over{{\rm tanh}(\Omega\rho_0)}}\Bigr)^\alpha
\biggr]^{\sigma/2}, \quad \alpha = \pm {{4 \lambda q}\over{\sigma}}
\label{dropg}
\end{equation} 
where $\rho_0$ is an
arbitrary constant. For $\alpha > 0$ the solution possesses physical
meaning at $\rho < \rho_0$ and becomes meaningless at $\rho > \rho_0$.
Outside the cylinder $\rho=\rho_0$ one can put $\vf \equiv 0$. This
trivial solution is stitched with the solution at $\rho\,=\,\rho_0$ under 
condition $\vf^{\p}(\rho_0) = 0$, which fulfills if and only
if $4 \lambda |q| > \sigma$. Consequently, at $\rho > \rho_0$ the Lagrangian 
becomes physically meaningless, however its limiting value at 
$\rho \to \rho_0 - 0$ is equal to zero. Continuing it at $\rho > \rho_0$,
one can consider the field be totally trivial in this area. Thus we get
the droplet-like configuration. The field $\vf$ steadily decreases 
from $\vf (0) = 1/\lambda$ to $\vf (\rho_0) = 0$ with
$\vf^{\p}(0) = 0$ (for $\sigma \ge 3$) and $\vf^{\p} 
(\rho_0) = 0$. The total energy of the "droplet" is defined as
\begin{equation}
E_f = {{4\pi q}\over{\lambda (\sigma - 1)}}
\end{equation} 
which remains unaltered even in flat space-time ($\Omega = 0$). 
It means that the "droplet" does not feel G$\ddot o$del gravitational field.
\subsection{Stability problem}
Let us now study the stability of the configuration obtained. In doing so
we consider the perturbed scalar field $\delta \vf = \chi(\rho, t)$ 
that leaves the cylindrical-symmetry of the system unbroken. The
linearized equation for the perturbed scalar field looks 
\begin{equation}
{{1}\over{\sqrt{-g}}}{{\pr}\over{\pr x^\mu}}\Bigl(\sqrt{-g}
g^{\mu\nu}\chi_{,\nu}\Bigr) + {{q^2}\over{2 (-g)}} P_{\vf\vf} \chi
= 0 \label{chi}
\end{equation}
Since, for the case considered, $g^{00} = (1 - {\rm sinh}^2 \Omega \rho)/
{\rm cosh}^2 \Omega \rho$, it is clear that the type of equation ~(\ref{chi})
changes on the surface where ${\rm sinh} \Omega \rho = 1$. The
Cauchi problem for (~\ref{chi}) is incorrect by Hadamard in the region, where
$g^{00} < 0$. In connection with this only the droplet-like 
solutions can be stable by Lyapunov, if they are located in the region 
where $g^{00} > 0$~\cite{BronIz}. 
Assuming that
\begin{equation}
\chi (\rho, t) = v (\rho) {\rm exp} (-i \varepsilon t)
\label{epsilon}
\end{equation}
from (~\ref{chi}) we get
\begin{equation}
v^{\p\p} + 2 \Omega {\rm coth} (2\Omega \rho) v^{\p} + 
\bigl[{{1 - {\rm sinh}^2 \Omega \rho}\over{{\rm cosh}^2 \Omega \rho}} 
\varepsilon^2 -
{{2 \Omega^2 q^2}\over{{\rm sinh}^2 (2 \Omega \rho)}} P_{\vf\vf}\bigr] v = 0
\label{v}
\end{equation}
The substitution
\begin{equation}
\eta (\zeta) = \xi (\rho) \cdot v (\rho) 
\label{subg}
\end{equation}
where
\begin{eqnarray}
\zeta = \int {{\sqrt{1- {\rm sinh}^2 \,\Omega \rho}}
\over{{\rm cosh}\, \Omega \rho}} d\rho, \quad
\xi = \bigl[ 4 {\rm sinh}^2\, \Omega\rho \bigl(1-{\rm sinh}^2\, \Omega\rho
\bigr)\bigr]^{1/4} \nonumber
\end{eqnarray}
leads the equation for perturbed field to the Liouville one
\begin{equation}
{{\pr^2 \eta}\over{\pr \zeta^2}} + \bigl[\varepsilon^2 - V]\,\eta = 0
\label{NFL}
\end{equation}
Here the effective potential $V (\vf)$ takes the form
\begin{eqnarray}
V(\vf) = {{\Omega^2}\over{(1 - {\rm sinh}^2 \Omega\rho)}} \Bigl\{
{{q^2 P_{\vf\vf}}
\over{2 {\rm sinh}^2 \Omega\rho}}
+{{(4 {\rm sinh}^6 \Omega\rho -16{\rm sinh}^4 \Omega\rho
+ 3 {\rm sinh}^2 \Omega\rho - 1)\,
{\rm cosh}^2 \Omega\rho}
\over{4 {\rm sinh}^2 \Omega\rho\,(1- {\rm sinh}^2 \Omega\rho )}}\Bigr\}
\nonumber
\end{eqnarray} 
For the interaction function, chosen in the form (~\ref{P}), one finds
$$P_{\vf \vf} = \lambda^2 \bigl(
{{2 \sigma^2 - 12 \sigma + 16}\over{\sigma^2 J^{4/\sigma}}} -
{{4 \sigma^2 - 12 \sigma + 8}\over{\sigma^2 J^{2/\sigma}}}  + 2 \bigr) $$
Taking into account that $\lim\limits_{\rho \to 0} J \to 1$ and
$\lim\limits_{\rho \to \rho_0} J \to 0$, it can be shown that
\begin{eqnarray}
\lim\limits_{\rho \to 0} V (\vf) \to & & + \infty \quad {\rm for} \quad
\sigma < 4 q \lambda \nonumber \\
\lim\limits_{\rho \to \rho_0} V (\vf) \to & & + \infty \quad {\rm for} \quad
\sigma \ge 5 \nonumber
\end{eqnarray}
Here we used the fact that $g^{00} > 0$, i.e., ${\rm sinh} \Omega \rho < |1|$.
Thus, as in the previous case we find that the droplet like configurations
are stable for $5 \le \sigma < 4 |q| \lambda$.  
\section{Conclusion}
\setcounter{equation}{0}
We obtained the regular particle-like solutions to the scalar 
field equations with induced nonlinearity in external gravitational
fields described by Freedman-Robertson-Walker and G$\ddot o$del Universes
respectively. Beside the usual solitons, a special type of regular
localized configurations, known as droplets, have been obtained.
It has been shown that the droplet-like configurations possess 
limited energy density and finite total energy and the droplets of 
different linear sizes up to the soliton share one and the same 
total energy. It is noteworthy to notice  that in the spherically
symmetrical case (i.e., in the FRW Universe) at $r_c\,\to\,\infty$ 
for $k\,=\,0$ droplet transfers to usual solitonian solution, while 
for $k\,=\,\pm 1$ this is not the case. It has also been shown that
in FRW space-time equations with different type of nonlinearities
may contain stable solutions, whereas in case of G$\ddot o$del 
Universe only the droplet-like configurations may be stable. It is
noteworthy to remark that in FRW space-time with $k = +1$ the 
field function is confined in the region $0\le r \le 1$ independent 
to the choice of interaction function $P (\vf)$.
 
\begin{figure}
\hspace{2cm}\epsffile{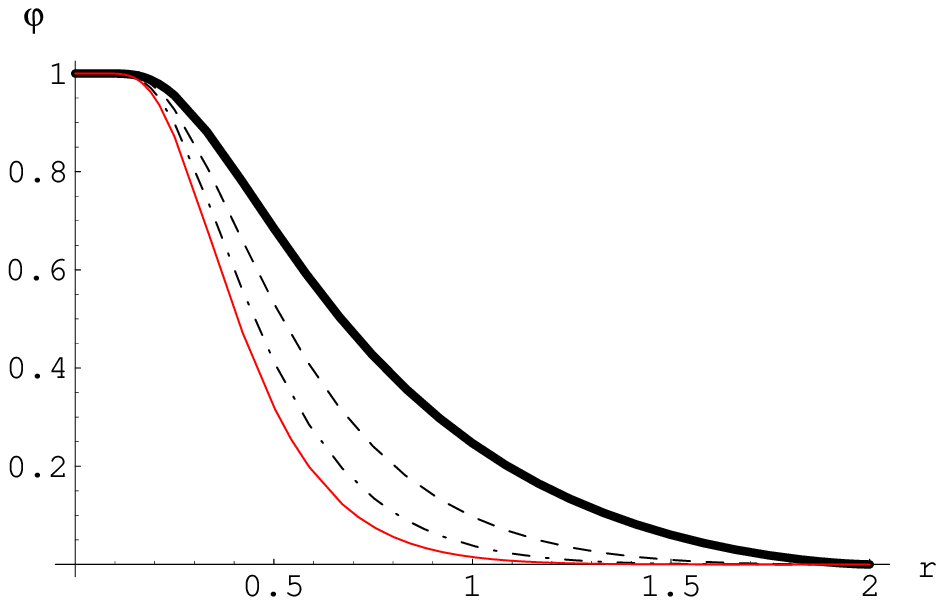}
\vspace{2cm}
\caption{Perspective view of droplet-like configurations for different values
of $\sigma$. Here the thick-line, dash-line, dash-dot-line and thin-line 
correspond to the value of $\sigma = 3,5,7,9$ respectively.}
\end{figure}
\newpage
\vskip 3cm
\begin{figure}
\hspace{2cm}\epsffile{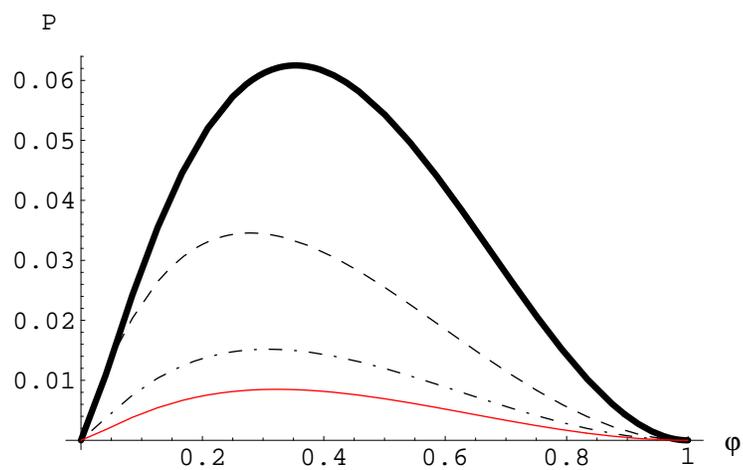}
\vspace{2cm}
\caption{Perspective view of interaction function $P(\vf)$ providing 
droplet-like configurations}
\end{figure}
\end{document}